\documentclass[usegraphicx,useAMS,usenatbib]{mn2e}

\title[A low-mass cluster of extremely red galaxies at z=1.10 in the GOODS Southern Field]
{A low mass cluster of extremely red galaxies at z=1.10 in the GOODS Southern Field}
\author[]{Anastasio D\'{\i}az--S\'anchez$^{1}$\thanks{E-mail:
andiaz@upct.es}, 
Isidro Villo--P\'erez$^{2}$, 
Antonio P\'erez--Garrido$^{1}$ 
\newauthor
and Rafael Rebolo$^{3,4}$\\
$^{1}$Dpto. F\'{\i}sica Aplicada, Universidad Polit\'ecnica de Cartagena, 
Campus Muralla del Mar, 30202 Cartagena, Murcia, Spain\\
$^{2}$Dpto. Electr\'onica, Tecnolog\'{\i}a de Computadores y Proyectos, 
Universidad Polit\'ecnica de Cartagena, Campus Muralla del Mar, \\ 
30202 Cartagena, Murcia, Spain\\
$^{3}$Instituto de Astrof\'{\i}sica de Canarias, V\'{\i}a L\'actea s/n, 38200 La Laguna, Tenerife, Spain \\
$^{4}$ Consejo Superior de Investigaciones Cient\'{\i}ficas, Spain}
\begin{document}


\pagerange{\pageref{firstpage}--\pageref{lastpage}} \pubyear{2006}

\maketitle

\label{firstpage}

\begin{abstract}
We have studied the spatial clustering of high redshift ($z>1$) extremely red objects (EROs) as a 
function of photometric redshift  in the GOODS Southern Field using public data. A remarkable overdensity 
of extremely red galaxies ($I-K_s > 4$) is found at an average photometric redshift $z_{\rm phot} =1.10$.
Nine objects (six are EROs)  within 50 arcsec of the brightest infrared galaxy in this overdensity  
present spectroscopic redshifts in the range $1.094<z_{\rm spec}< 1.101$ with  
a line-of-sight velocity dispersion of $\sigma_v=433^{+152}_{-74}$ km s$^{-1}$ typical of an Abell richness 
class $R=0$ cluster.   Other potential members of this cluster,  designated as
GCL J0332.2-2752, have  been identified using photometric redshifts and the  
galaxy density profile  studied  as a function of radius. The mass of the cluster
 is preliminary estimated at M$_{cl}\sim 5-7 \times 10^{13} M_{\sun}$. Using available Chandra data, 
we limit the rest-frame  X-ray luminosity  of the cluster to less 
than $L_X=7.3 \times 10^{42}$ erg s$^{-1}$ (0.5-2.0 keV). 
This low-mass, low L$_X$ cluster at  $z>1$ shows the potential of EROs to trace clusters 
of galaxies at high redshift.

\end{abstract}


\section{Introduction}

Clusters of galaxies trace the largest gravitationally bound mass concentrations in the Universe.
The number density of galaxy clusters as a function of mass and redshift provide a testbed for 
cosmological models and set constraints on cosmological parameters such as the density of
dark matter and the equation of state of dark energy \citep{CC04,MP04}. Clusters at various
redshifts also provide samples of galaxies in dense environments to confront galaxy formation 
and evolution theories.

Optical and near infrared follow-up of X-ray sources is a proven  technique for constructing large 
samples of distant clusters \citep{Ro02}, however it has provided only about ten clusters with X-ray emission 
confirmed at $z>1$, the highest redshift cluster being at $z=1.41-1.45$ \citep{MR05,St06}.  
The majority of the other confirmed clusters at $z>1$ have been identified in photometric surveys 
around radio-sources \citep{BG03,Bs00,N01}. 

High-redshift low X-ray luminosity (low-T) clusters are often below the limit of X-ray surveys.
 Deep infrared surveys offer an alternative path to identify 
high redshift galaxy clusters. The cores of these clusters  are expected to be dominated by elliptical 
galaxies which at $z>1$  appear as extremely red objects 
(EROs) with colours $I-K>4$. A significant fraction of these red objects ($30-45 \% $ of the total number at 
$K_s=20-22$ in the general field) are early-type galaxies and have redshift distributions with 
$z_{\rm mean}\approx 1.2$ \citep{MC04,SE97}. Several spectroscopic studies have shown the EROs population 
is heterogeneous, being mainly formed by old passively evolving distant elliptical galaxies and 
extremely dust reddened starburst galaxies \citep{C02,C03}. EROs have strong spatial clustering, 
comparable to that of present-day luminous ellipticals, which has been interpreted as evidence that 
distant EROs and nearby ellipticals are evolutionary linked, i. e., EROs could be the progenitors of 
present day massive elliptical galaxies. At least two distant clusters of EROs have been confirmed 
in general field surveys  at $z=1.27$ \citep{SE97} and $z=1.41$ \citep{St05}, respectively. 

In this paper we study the clustering of high redshift ($z>1$) EROS as a function of photometric redshift in the
GOODS Southern field and report on  a cluster of galaxies at $z=1.10$ with
at least ten spectroscopically confirmed members within a 92 arcsec radius region.
The use of public data, catalogues, and the cluster-finding algorithm are described in Section 2. 
The basic properties of this galaxy cluster are described in Section 3 and conclusions 
are given in Section 4. We assume a standard concordance cosmological model throughout, 
with $H_0=70$ km s$^{-1}$ Mpc$^{-1}$, $\Omega_m=0.3$, and $\Omega_\Lambda=0.7$.

\begin{table*}
 \centering
 \begin{minipage}{140mm}
  \caption{ERO overdensities at $1 < z_{\rm phot} < 2$.}
 \begin{tabular}{@{}cccccccc@{}}
 \hline
    $z_{\rm cl}$ & Photometric & Number & EROs with & 
    Overdensity & Total ERO density \footnote{Density of EROs with $z_{\rm phot}$ within the photometric redshift
    interval in the whole GOODS-MUSIC dataset.} &RA \footnote{Right ascension of the centroid of the coordinates 
    for the EROs, in decimal degrees (J2000).} & 
    Dec \footnote{Declination of the centroid of the coordinates for the EROs, in decimal degrees (J2000).} \\
     & redshift interval  & of EROs &  measured $z_{spec}$& 
    factor & (galaxies/arcmin$^2$) &  J2000 &  J2000 \\
 \hline \hline
 1.00 & 0.940--1.060 & 10 & 4 & 3.78 & 0.33 & 53.09576 & -27.87832\\
 1.10 & 1.034--1.166 & 17 & 10 & 4.11 & 0.51& 53.08198 & -27.88069\\
 1.70 & 1.598--1.802 & 15 & 1 & 3.10 & 0.60 & 53.11302 & -27.71836\\
 1.95 & 1.833--2.067 & 13 & 0 & 3.10 & 0.52 & 53.06804 & -27.83184\\
\hline
\end{tabular}
\end{minipage}
\end{table*}

\section[]{Sample and cluster-finding algorithm}

We have studied the spatial distribution of EROS (defined here as objects with $i-K_s >4$). 
in the GOODS-MUSIC (MUltiwavelength Southern Infrared Catalogue) dataset \citep{Gr06} as a 
method to identify clusters of galaxies at high redshift. The GOODS-MUSIC catalogue includes 
the $F435W$ ($B$), $F606W$ ($V$), $F775W$ ($i$), and $F850LP$ ($z$) ACS data, the $JHK_s$ VLT-ISAAC data, 
the Spitzer data provided by IRAC, and the available $U$-band data from 2.2ESO and VLT-VIMOS. 
The images from space and ground-based telescopes are of different resolution and depth and the
colours are PSF matched.
In this work we use the Vega system while in  the GOODS-MUSIC catalogue magnitudes are given in the AB system. 
The ACS data and the VLT-ISAAC data were converted to the Vega system using 
$(B,V,i,z,J,H,K_s)_{\rm Vega}=$$(B,V,i,z,J,H,K_s)_{\rm AB}+$ $(0.106,-0.088,-0.398,-0.536,-0.90,-1.38,-1.86)$
\footnote{http://www.eso.org/science/goods/release/20050930/ and http://www.stsci.edu/hst/acs/analysis/zeropoints/}. 
The catalogue is complete at $i_{\rm comp}=26.1$ and $K_{s,{\rm comp}}=22$ (Vega scale) \citep{Gr06} and 
lists photometric redshifts $z_{\rm phot}$ for all the objects with  average uncertainty  $\sigma_{\rm phot}=0.06$. 

The spectroscopic catalogues used by  GOODS-MUSIC  are the COMBO-17 survey \citep{Wo01}, 
the CXO-CDFS survey \citep{Sz04}, the K20 survey \citep{Mi05}, the GOODS V1.0 survey \citep{Va05}, 
the VVDS survey \citep{LF04} and the MASTER catalogue 
\footnote{http://www.eso.org/science/goods/spectroscopy/CDFS\_Mastercat/}.
Spectroscopic redshift $z_{\rm spec}$ 
are available for more than 1000 objects. We will also work with the GOODS V2.0 survey \citep{Va06}. 
As early-type galaxies with $1\le z\le 2$ have their rest-frame light peaks at the near-infrared band 
we only consider galaxies whith $K_s \le K_{s,{\rm comp}}$, where $K_{s,{\rm comp}}$ is
the completeness limit of the catalogue.
  
A simple cluster-finding algorithm was implemented to  search for overdensities of EROs in the GOODS-MUSIC dataset 
with redshift in the range $1< z_{\rm phot} < 2$. We adopted 0.8 Mpc as the radius of the circles used in the density 
determination ($\sim 1.6$ arcmin for $1<z<2$ according to the  concordance
cosmological model). This is smaller than  the Abell radius of a typical cluster $r_{\rm Abell}\sim1.5$ Mpc but
we expected that  EROs mostly concentrate in the core of the clusters. 

To detect spatial clustering at a given redshift  $z$ we first selected the total sample of EROs in the catalogue 
with $z_{\rm phot}$ within the interval $[z-\sigma_{\rm phot}z;z+\sigma_{\rm phot}z]$, where 
$\sigma_{\rm phot}$ is the average uncertainty of photometric redshifts. The density of EROs was then computed
in circles of radius $\sim 1.6$ arcmin centred around each ERO of the sample, and this value was
compared with the density of  EROs in the whole GOODS Southern survey ($\sim $ 140 arcmin$^2$) at the 
given redshift. Only overdensities  a factor 3 higher than the average were considered as potential 
galaxy clusters, provided the objects also followed a galactic red sequence in the $i-K_s$ vs. $K_s$ 
diagramme. The redshift of the candidate cluster $z_{\rm cl}$ was then preliminary adopted as the average value 
of the photometric redshifts of the EROs in the overdensity. 

We found four ERO overdensities with main properties listed in Table 1. For each potential cluster we 
checked the availability of spectroscopic redshifts. 
The overdensity at $z_{\rm cl}=1.00$ contains four objects with  spectroscopic
redshift $z_{spec}$ but they lie in an interval $\Delta z_{spec}=z_{max}-z_{min}=0.08$ which is too 
wide for a cluster of galaxies. For the second overdensity listed in Table 1, we find 10 EROs with 
spectroscopic redshifts  within the indicated redshift interval, six lie in a $\Delta z_{spec}=0.004$ 
interval around $z_{\rm mean}=1.098$, confirming that this region contains an excellent galaxy cluster 
candidate. Of the remaining four objects, two are at $z=1.12$ and the other at $z=1.044$ making 
unplausible a physical relation to the galaxy cluster. 
We note that the  two remaining higher redshift candidate overdensities in Table 1 are less reliable given 
the wide redshift interval under consideration. Unfortunately, there is insufficient spectroscopic 
information and we shall await for additional data before making any further consideration on these
potentially interesting overdensities.

\begin{table*}
 \centering
 \begin{minipage}{140mm}
  \caption{Spectroscopic candidates of the GCl J0332.2-2752 cluster of galaxies at $z=1.10$.}
 \begin{tabular}{@{}lllllllll@{}}
 \hline
   \ \ ID & \ \ \ \ RA \footnote{Right ascension, in decimal degrees (J2000).} &
   \  \ \ \ Dec \footnote{Declination, in decimal degrees (J2000).} &
  \ \ \ $z$ \footnote{Spectroscopic redshift.}&
    $qz$ \footnote{Quality of spectroscopic redshift (0 = very good, 1 = good, 2 = uncertain, 3 = bad quality).} &
   \ \ \ class \footnote{Spectroscopic class \citep{Gr06}.} &
   \ \ catalogue \footnote{Reference spectroscopic catalogue.} &  \ \ \ $K_s$  & \ \ \  $i-K_s$ \\

    &\ \ \  J2000 &\ \ \  J2000 & &  &  & &  \\

 \hline \hline
 2859& 53.09842&-27.88740& 1.103&  0& EMISSION &GOODSV2.0& 21.83 &2.35 $\pm 0.10$ \\
 3185& 53.06122&-27.88284& 1.101&  0& EMISSION &GOODSV2.0 & 21.22 &2.38 $\pm 0.05 $ \\
 3315& 53.07283&-27.88000& 1.095&  0&EARLY    &GOODSV2.0 & 18.06 &4.41 $\pm 0.04 $ \\
 3490& 53.06739&-27.87814& 1.094&  1&EARLY    &GOODSV2.0 & 19.04 &4.18  $\pm 0.06 $ \\
 3586& 53.06193&-27.87676& 1.101&  1&GALAXY   &VVDS   & 20.05 &2.67   $\pm 0.04 $ \\
 3619& 53.07276&-27.87632& 1.100&  0&EARLY    &GOODSV2.0 & 18.21 &4.25  $\pm 0.04 $ \\
 3656& 53.06210&-27.87654& 1.101& 2&GALAXY   &MASTER & 22.19 &2.27   $\pm 0.10 $ \\
 3698& 53.07340&-27.87458& 1.098&  0&COMPOSITE&GOODSV2.0 & 17.52 &3.99 $\pm 0.02 $ \\
 3920& 53.07155&-27.87245& 1.097&  0&AGN      &CXO-CDFS & 18.07&4.18  $\pm 0.04 $ \\
 3941& 53.08044&-27.87204& 1.096&  0&EARLY    &GOODSV2.0 & 18.12 &4.60 $\pm  0.07 $ \\
\hline
\end{tabular}
\end{minipage}
\end{table*}

\section[]{The GCl J0332.2-2752 Cluster of Galaxies at $z=1.10$}

\begin{figure}
\includegraphics[width=87mm]{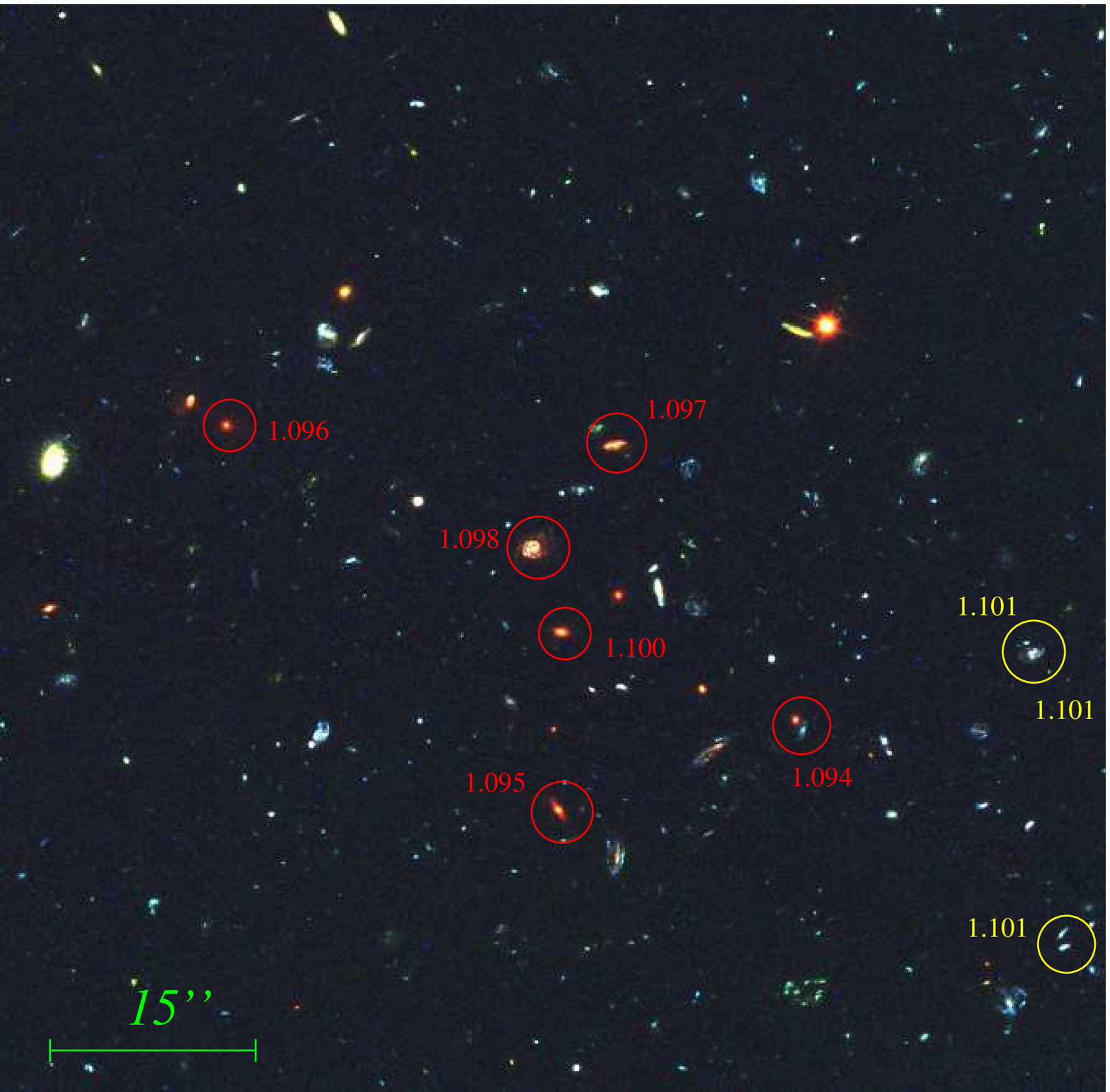}
\caption{ 
$BVz$ false-colour ACS (Hubble) image of the GCl J0332.2-2752 cluster of galaxies covering 1.36 arcmin on a side. 
The EROs members of the cluster with measured spectroscopic redshift are marked with red open circles and their
$z_{\rm spec}$ are given by red numbers. The no-ERO members with measured spectroscopic redshift are marked with
yellow open circles and their $z_{\rm spec}$ are given by yellow numbers. 
The galaxy with $z=1.098$ is the most luminous, the AGN with  $z=1.097$  is of type QSO-2.
The double redshift label for one circle indicates independent redshifts for two close galaxies.
North is up and East is left.
\label{fig1}}
\end{figure}

We now concentrate on the EROs overdensity at $z_{\rm cl}=1.10$. In Table 2 we list identification numbers, 
coordinates, redshifts, spectroscopic class, spectroscopic catalogue, $K_s$ magnitude and $i-K_s$ colour for 
ten galaxies in this region with spectroscopic redshifts in the range $1.094<z_{\rm spec}< 1.103$. 
From these ten galaxies we obtain a  mean spectroscopic redshift for the cluster  $<z>=1.0986$ 
and a rms dispersion $\sigma = 0.0028$. We tentatively determine the equatorial coordinates of the cluster centre
as the centroid of the coordinates for the six EROs in Table 2, which are  $\alpha_{J2000.0}= 03^h 32^m 17.5^s$, 
$\delta_{J2000.0}=-27^{\degr} 52^m 32^s$. Hereinafter we will refer to this cluster as GCl J0332.2-2752.  
In Fig. \ref{fig1} we plot a $BVz$ false-colour image, from the ACS images, of the region showing 
the location of nine of these objects within 50 arcsec of the brightest infrared galaxy at z=1.098 which is close to 
the centre of the cluster. Six of them are EROs (marked in red in the figure) and lie within a radius of 
24 arcsec from this galaxy. The other four are no-ERO galaxies, three appear in the figure marked with 
yellow open circles. The galaxy ID 2859 with the highest redshift 
($z_{\rm spec}=1.103$) is also at the largest angular distance 92 arcsec of the central region of the cluster
and it is not shown in the image. We also note the presence of an AGN ( $z=1.097$) of type QSO-2 near the 
central region \citep{SB04}. The nine galaxies in the image are classified as follows: six are EROs, 
out of which four belong to the ``early'' class, the most luminous galaxy is a ``composite'' 
(early+late type galaxy) class and one is an AGN. Out of the three no-EROs, two belong to the 
class ``galaxy'' (this means that there is no information in the catalogue about what type of 
galaxy it is) and the other to ``emission'' class, as defined by \cite{Gr06}.

We now estimate the velocity dispersion of the GCl J0332.2-2752  along the line-of-sight 
from the ten galaxy members with available spectroscopic redshifts obtaining 
$\sigma_v=433^{+152}_{-74}$ km s$^{-1}$, with 
the 68\% confidence uncertainty estimated  according to the expression given by \cite{DD80}. 
This velocity dispersion corresponds to the median value for Abell richness class $R=0$ \citep{YE03}. 
In order to estimate the mass and the size of the cluster we calculate
the radius $R_{200}$, which approximates the virial radius. It is the radius inside which the
density is 200 times the critical density, $200 \rho_c(z)=M_{\rm cl}/(4\pi/3)R^3_{200}$. Using the
redshift dependence of the critical density and the virial mass, $M_{\rm cl}=3\sigma_v^2R_{200}/G$,
we obtain \citep{FZ05}:
\begin{equation}
R_{200}=2.47\frac{\sigma_v}{1000 {\rm \ km \ s^{-1}}}\frac{1}{\sqrt{\Omega_m(1+z)^3+\Omega_\Lambda}} 
\ {\rm Mpc} \;.
\end{equation}
For the cluster GCl J0332.2-2752 we have $R_{200}=0.57^{+0.2}_{-0.1}$ Mpc or $R_{200}=69^{+24}_{-11}$ arcsec. 
We combine the virial mass with the expression for $R_{200}$ to obtain an estimate of the cluster mass in 
terms of $\sigma_v$ and cosmological parameters:
\begin{equation}
M_{\rm cl}=1.71\times 10^{15}\left(\frac{\sigma_v}{1000 {\rm \ km \ s^{-1}}}\right)^3 
\frac{1}{\sqrt{\Omega_m(1+z)^3+\Omega_\Lambda}} \ M_{\sun} \;,
\end{equation}
which gives a mass for our cluster $M_{\rm cl}=7^{+2}_{-1}\times 10^{13} M_{\sun}$.
If we use the more empirical $M_{200}-\sigma_v$ relation given by expression 47 in \cite{Vo05} we have 
$M_{\rm cl}=4.9^{+1.6}_{-0.8}\times 10^{13} M_{\sun}$. 
If the intracluster gas shares the same dynamics of typical galaxy clusters its temperature $T$ can be estimated 
from $k_BT\simeq \mu m_p \sigma^2_v \simeq 6\frac{\sigma^2_v}{10^6 ({\rm km\ s^{-1}})^2}$ keV \citep{Ro02}, 
where $\mu$ is the mean molecular weight ($\mu\sim 0.6$ for a primordial contribution with a $76\%$ fraction 
contributed by hydrogen) and $m_p$ is the proton mass. We estimate $k_BT\simeq 1.1^{+0.8}_{-0.4}$ keV.
This compares well with the value, $k_BT\simeq 1.5 \pm 0.6 $ keV, inferred from the $\sigma_v$-T 
prescriptions by \cite{LB93}.

\begin{figure}
\includegraphics[width=80mm]{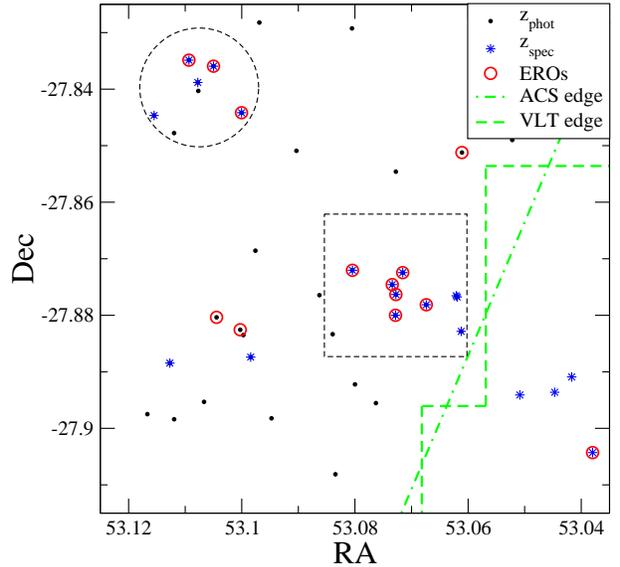}
\caption{Distribution of galaxies with $z_{\rm phot}$ within the interval
$[1.088;1.107]$ in the region of GCl J0332.2-2752 . The edges of the ACS images and VLT-ISAAC images are shown.
Photometric candidates are marked by dots, galaxies with measured 
$z_{\rm spec}$ are marked by blue stars and EROs are also marked by
red open circles. The field of Fig \ref{fig1} is also shown as a dashed box.
A group of galaxies is shown within the dashed circle.
There are also four galaxies in the bottom-right part, which are outside of the GOODS-MUSIC area,
with measured spectroscopic redshift. 
\label{fig2}}
\end{figure}

\begin{figure}
\includegraphics[width=82mm]{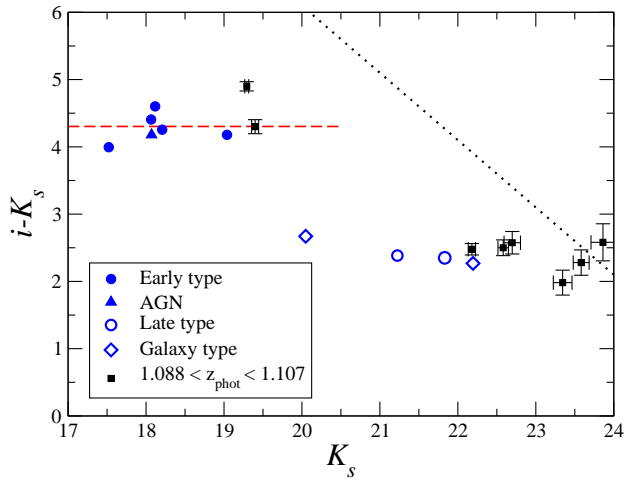}
\caption{Colour-magnitude diagram for the candidates to the GCl J0332.2-2752 cluster of galaxies with $z_{\rm phot}$ 
within the interval $[1.088;1.107]$. The members of the clusters with measured $z_{\rm spec}$ are marked 
by blue colours and their errors are equal or smaller than the sizes of the symbols. 
The red dashed line is the mean value of the $i-K_s$ colour for the EROs and the 
dotted line is the completeness $i$-band limit.
Colours are PSF matched between the ACS and ISAAC magnitudes and they are in the Vega system.
\label{fig3}}
\end{figure}

A comparison of photometric and spectroscopic redshifts
for objects around z=1.1 in the GOODS-MUSIC sample shows little scatter suggesting that the uncertainties
of the photometric redshifts are lower than 1 \%. This implies  that new cluster members can be reliably identified 
using photometric redshifts. In Fig. \ref{fig2} we plot the spatial distribution of galaxies with 
redshifts $1.088 < z < 1.107$ 
($<z> \pm 3\sigma$) in a wider region of the sky. Both, objects with spectroscopic and photometric redshift 
determinations are plotted. A number of potential cluster members are found in the immediate vicinity 
of the region that contains the higher ERO overdensity. A colour-magnitude diagram, $i-K_s$ vs. $K_s$
of galaxies within a radius of 1.6 arcmin of the cluster centre is shown in Fig. \ref{fig3}. 
We can see that the most luminous EROs follow a well defined red sequence, as expected. We also
find at fainter magnitudes $K_s>$20 a  possible sequence of galaxies with average colour $i-K_s=2.5$. It appears
that both sequences are similarly populated and that the fainter one is mostly due to 
spiral galaxies \citep{BO84,TS03}. Following \cite{HM05} we can estimate the richness for the cluster 
as the number of galaxies in the upper  red sequence, we find here $N_{\rm gal_r}=8$. 
Low redshift galaxy clusters with this value show velocity dispersions which are similar \citep{HM05} within
error bars to the one measured for GCl J0332.2-2752.
  
\begin{figure}
\includegraphics[width=82mm]{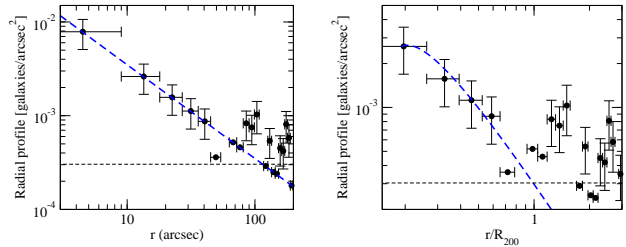}
\caption{
The galaxy density profile as a function of the radius for the GCl J0332.2-2752 cluster of galaxies, here we 
consider galaxies with $z_{\rm phot}$ within the interval $[1.088;1.107]$. The black dashed lines in both 
figures are the mean density of galaxies within the interval $[1.088;1.107]$ in the whole GOODS-MUSIC 
area which is 0.000303 galaxies/arcsec$^2$.
Left: the blue dashed line is a fit of a power law $\Sigma = N_0 r^{-\alpha}$ with exponent 
$\alpha=1.0 \pm 0.1$ and $N_0=0.04 \pm 0.01$. 
Right: the galaxy density profile as a function of the radius divided by $R_{200}$ for the GCl J0332.2-2752 
cluster of galaxies. The best-fitting NFW profile is shown by the blue dashed line.
\label{fig4}}
\end{figure}

In Fig. \ref{fig4} we give the galaxy density profile of the cluster as a function of radius, here
we consider galaxies with $z_{\rm phot}$ within the interval $[1.088;1.107]$.
We evaluate the density profile taking annuli centred at the optical centre of the cluster, when the radius 
arrives to the edge of the field images ($\sim 55$ arcsec) we take semiannuli in the direction away 
from the edges (east direction). The dashed line in  Fig. \ref{fig4} left is a fit with a power law 
$\Sigma = N_0 r^{-\alpha}$ with exponent $\alpha=1.0 \pm 0.1$ and $N_0=0.04 \pm 0.01$, which agrees
with the behaviour found in  low redshift clusters \citep{HM05}. 
We now fit the density profile with a NFW model \citep{NF93}, where the 3D number density
profile is $n(x)=n_0 x^{-1}(1+x)^{-2}$ with $n_0$ the normalisation and $x=r/r_s$, where 
$r_s$ is a scale radius. The $r_s$ scale radius is related by $r_s\equiv R_{200}/c_{g}$ to the 
virial radius $R_{200}$ and the galaxy concentration parameter $c_g$. 
The surface density is then an integral of the 3D profile 
$\Sigma (x)= 2n_0 r_s \int_0^{2/\pi} \cos\theta(\cos \theta+x)^{-2} d\theta$ 
\citep{Ba96}. Using $R_{200}=0.69$ arcsec we fit the data in the right panel of Fig. \ref{fig4} to
obtain $n_0=0.11 \pm 0.02$ galaxies/arcsec$^2$ and $c_g= 15 \pm 4$. 
The galaxy concentration parameter is higher than those found for low redshift clusters 
$c_g\sim 2-7$ \citep{HM05,PA05} and in simulations $c_g\sim 2-10$ \citep{Ji00}. The excess
of galaxies that can be noticed at $r/R_{200}=2.45$ in the right panel of
Fig. \ref{fig4} corresponds to a group of galaxies located at the top of Fig. 2 (dashed circle). 
There are spectroscopic redshifts for five of these galaxies (of which 3 are EROs). The mean  redshift of
the group  is $<z>=1.0964$,  remarkably similar to that of GCl J0332.2-2752 and the rms dispersion is 
$\sigma=0.0020$. The centroid of this group in equatorial coordinates is $\alpha_{J2000.0}= 03^h 32^m 25.8^s$,
$\delta_{J2000.0}=-27^{\degr} 50^m 22^s$, thus it is located at an angular distance of 169 arcsec (1.41 Mpc) 
from the cluster. It is then  plausible that both systems of galaxies form part or will
become part of the same cluster.
There are also four galaxies in the bottom-right part of Fig. 2, which are outside of the GOODS-MUSIC area, 
with spectroscopic redshift measured in \cite{LF04}. The mean redshift of these galaxies is $<z>=1.099$ and 
the rms dispersion is $\sigma=0.0018$, the nearest galaxy to the centre of GCl J0332.2-2752 is 
at 102 arcsec (0.84 Mpc) and the farthest is an ERO at 156 arcsec (1.28 Mpc).
It is unfortunate that the location of GCl J0332.2-2752 near the edge of the GOODS field prevents 
identification of other potential substructures. According to hierarchical structure formation it is 
plausible that distant clusters  may be conformed by smaller substructures with rather 
cool gas, e.g. \cite{FE96}.
In the Chandra Deep Field South (CDF-S) there is evidence for sheets in the distribution of galaxies at 
$z=0.666$, $z=0.734$, $z=1.096$, $z=1.221$, $z=1.300$ and $z=1.614$ \citep{Va06}. 
Two X-ray clusters have been identified in correspondence with the spikes in the redshift distribution 
at $z=0.666$, $z=0.734$ \citep{GZ02,Gi04}, these clusters are knots of sheet-like structures extending 
over several Mpc. Using a friend-of-friend algorithm, \cite{AM05}  also 
identified a compact structure of galaxies called ``Structure 15''  at z=1.098 in the location of 
our proposed cluster. These authors argued that there is no clear red sequence of galaxies associated with this 
structure. However, we have shown in the $i$ vs $i-K_s$ plot of Fig. 3 that indeed there is a red sequence of 
galaxies suggestive of a cluster and as we can see in Fig. 4 that the galaxy density profile is also typical of a 
cluster. The existence of potential substructures deserve further investigation as it may indicate the cluster 
is in the process of formation.  

\subsection{Morphology}

\begin{figure}
\includegraphics[width=125mm]{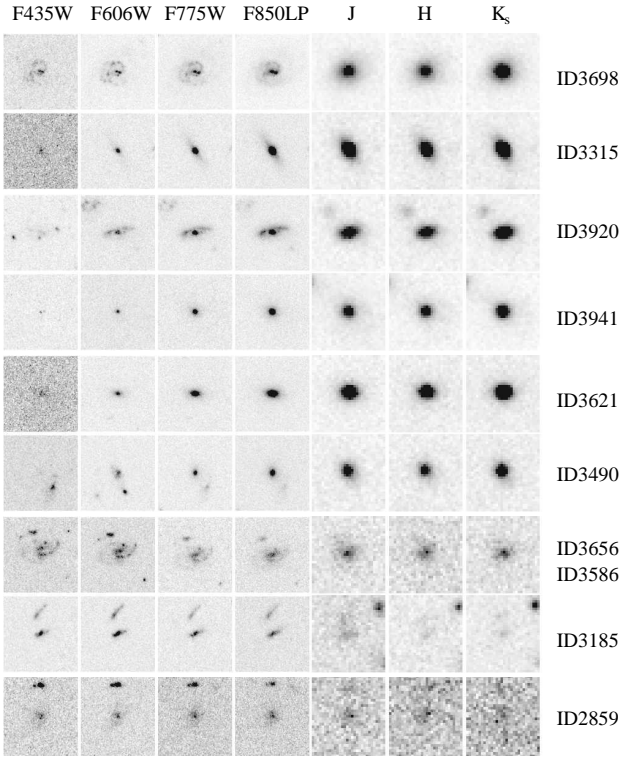}
\caption{Multiband ACS and VLT-ISAAC image of the galaxies belonging to the GCl J0332.2-2752 cluster of galaxies. 
Each ACS image is 3.6 arcsec on a side and each VLT-ISAAC image is 4.5 arcsec on a side.
\label{fig5}}
\end{figure}

We discuss in the following paragraph the optical/near infrared morphological appearance of the galaxies with 
confirmed spectroscopic redshift in the GCl J0332.2-2752 cluster.
In Fig. \ref{fig5} we show the available ACS and VLT-ISAAC images for these galaxies.
The six EROs look very compact and regular in the VLT-ISAAC images, nevertheless ID3698 has a double bulge
and spiral arms in the ACS images, perhaps due to merging or cannibalism, and ID3920 is an AGN. 
The other four EROs are visually classified as E/S0 systems. 
ID3586 is an SBb or SBc galaxy which is merging with the small galaxy ID3656 (left up in the ACS images). 
ID3185 is too faint to be classified but it would be an Sa or Sb system, the galaxy which is up in the 
images has a higher photometric redshift. The last object ID2859 is also too faint to be classified 
but look like a spiral galaxy, up in the images there is a galaxy with a higher photometric redshift.

\subsection{X-Ray data}

\begin{figure}
\includegraphics[width=41mm]{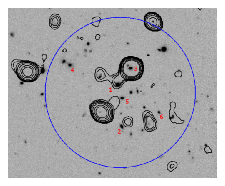}
\includegraphics[width=41mm]{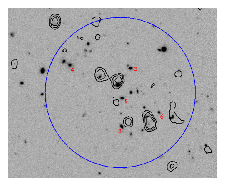}
\caption{
Left: Chandra contours of X-ray emission (0.5-2.0 keV) in  GCl J0332.2-2752 overlaid on the  VLT-$K_s$ image.
The Chandra data are  smoothed with a 5 arcsec FWHM Gaussian.  Contours correspond to $[2,3,5,7,15,25,50]\sigma$ 
above the local background. The blue circle has a radius of 35 arcsec and it is approximately centred in the 
position of the optical cluster. Right: the same but after applying a 
PSF subtraction of the brightest sources. 
The EROs members of the cluster are marked in both figures 
by a number from the brightest to lowest luminosity EROs (1 is the brightest) in the $K_s$ band. 1 is ID3698, 2 (ID3315), 
3 (ID3920), 4 (ID3941), 5 (ID3619), and 6 (ID3490).
\label{fig6}}
\end{figure}

We use the X-ray images of the CDF-S in order to 
detect possible diffuse or extended emission from the GCl J0332.2-2752 cluster of galaxies \citep{RT02}.
No X-ray extended sources have been found  in 
the region where  the cluster is located \citep{GZ02}. We work with the image of the soft band (0.5-2.0 keV events) 
taken from the CDF-S web page \footnote{http://www.mpe.mpg.de/\~{}mainieri/cdfs\_pub/index.html}.
In Fig. \ref{fig6} left we show the VLT-$K_s$ image of the cluster of galaxies with overlaid 
Chandra X-ray contours, data are from the 0.5-2.0 keV events smoothed with a 5 arcsec FWHM Gaussian. 
Here we can see the X-ray emission from the AGN is weakly extended towards the brightest galaxy in the cluster. 
Now we evaluate the net counts in the 0.5-2.0 keV band in a 35 arcsec aperture of the X-ray emission 
centred in the optical centroid of the cluster.  
We take a 35 arcsec aperture in order to compare  with the low luminosity clusters  in \cite{SH01}. 
We find  1067 counts with an effective exposure time of 777 ks (calculated from the exposure map given in \cite{GZ02}).
The background was estimated locally using three source-free circular apertures (r=10 arcsec) located 
around the cluster. The background corrected counts in the r=35 arcsec aperture was $319 \pm 30 $ 
which corresponds to a flux of $(1.9\pm 0.2) \times 10^{-15}$ erg cm$^{-2}$ s$^{-1}$ and a rest frame luminosity of 
$L_X=(1.6 \pm 0.2) \times 10^{43}$ erg s$^{-1}$ (0.5-2.0 keV). 

In order to study a possible underlying extended source we subtract the point-spread function (PSF) of the 
brightest sources near to the core cluster from the soft band X-ray smoothed image. 
The result is plotted in Fig. \ref{fig6} right which  shows the VLT-$K_s$ image 
with the X-ray contours  from the PSF subtracted X-ray image. The residual X-ray contours are located close
but do not coincide precisely with the positions of galaxies. 
The centre of the X-ray emission within the 35 arcsec aperture (denoted in the figure as a blue circle),
was calculated as the flux-weighted centroid of all events in this image and  agrees well with the optical 
centre of the cluster. Nevertheless since there is no obvious extended emission, 
using the PSF subtracted X-ray image, we simply evaluate the flux in the aperture by integrating 
all the counts. After correction for background, it results $153 \pm 30$ net counts which leads to 
an upper limit to the possible X-ray flux  emission in the band of 
$(9 \pm 2) \times 10^{-16}$ erg cm$^{-2}$ s$^{-1}$. This flux corresponds to
a rest frame luminosity of $L_X=(6.0 \pm 1.3) \times 10^{42}$ erg s$^{-1}$. Thus, we limit the X-ray luminosity
of the cluster to less than $L_X=7.3 \times 10^{42}$ erg s$^{-1}$ (0.5-2.0 keV). Such low value is consistent with
expectations  from the $L_X-\sigma_v$ relationship \citep{Mu00}. GCl J0332.2-2752 is the less luminous 
high redshift ($z>1$) galaxy cluster found so far (see e.g. \cite{BV06,SH01}).

\section{Conclusions}

We have used a cluster-finding algorithm based in the clustering of EROs at a given photometric redshift
in order to find high redshift galaxy clusters in the GOODS-MUSIC dataset. We identify  a cluster 
of distant galaxies, GCl J0332.2-2752, at $z=1.10$ in the GOODS Southern Field. Nine galaxies present 
spectroscopic redshift in the range of $1.094<z_{\rm spec}< 1.101$ and within 50 arcsec (0.42 Mpc) 
from the brightest infrared galaxy. Six of them are EROs and lie within a radius of 24 arcsec (0.20 Mpc) 
from this galaxy.
The velocity dispersion of the GCl J0332.2-2752 cluster of galaxies along the line-of-sight is 
$\sigma_v=433^{+152}_{-74}$ km s$^{-1}$ which corresponds to the Abell richness class $R=0$, 
virial radius $R_{200}=0.6^{+0.2}_{-0.1}$ Mpc, mass $M_{\rm cl}=4.9^{+1.6}_{-0.8}\times 10^{13} M_{\sun}$
and intracluster gas temperature $k_BT \sim 1.5 $ KeV. We limit the X-ray luminosity of the cluster to 
less than $L_X=7.3 \times 10^{42}$ erg s$^{-1}$ (0.5-2.0 keV). This upper limit  is  consistent with the
value predicted from the local $L_X-\sigma_v$ relation. 

This is one of the lower mass clusters found to date at redshift $z>1$. Such clusters 
will be hard to find with any other method: in X-ray it is difficult even with very deep exposures;  
they may  be out of reach for the upcoming generation of Sunyaev-Zeldovich surveys; using weak lensing 
techniques it is very complicated to detect clusters at these low masses even with deep HST observations 
(detection of groups was achieved via stacking of shear signals). Thus, beside studying the environment 
of high-$z$ radio sources, deep optical/near infrared surveys appear as a valid alternative to study the 
population of low-mass galaxy clusters at $z>1$.

\section*{Acknowledgments}
This work was supported by Project No. 03065/PI/05 from the Fundaci\'on S\'eneca and Plan Nacional 
de Astronom\'{\i}a y Astrof\'{\i}sica Project AYA2005-06543.
We acknowledge to the ESO/GOODS and the Chandra Deep Field projects and the teams that carried out programs
LP168.A-0485 and LP170.A-0788 with the  Very Large Telescope at the ESO Paranal Observatory.

\bsp

\label{lastpage}

\end{document}